# Is Dynamic Rumor Detection on social media Viable? An Unsupervised Perspective

Chahat Raj[1], Priyanka Meel[2]
[1]chahatraj58@gmail.com, [2]priyankameel86@gmail.com
Department of Information Technology, Delhi Technological University, Delhi, India

**Abstract**: With the growing popularity and ease of access to the internet, the problem of online rumors is escalating. People are relying on social media to gain information readily but fall prey to false information. There is a lack of credibility assessment techniques for online posts to identify rumors as soon as they arrive. Existing studies have formulated several mechanisms to combat online rumors by developing machine learning and deep learning algorithms. The literature so far provides supervised frameworks for rumor classification that rely on huge training datasets. However, in the online scenario where supervised learning is exigent, dynamic rumor identification becomes difficult. Early detection of online rumors is a challenging task, and studies relating to them are relatively few. It is the need of the hour to identify rumors as soon as they appear online. This work proposes a novel framework for unsupervised rumor detection that relies on an online post's content and social features using state-of-the-art clustering techniques. The proposed architecture outperforms several existing baselines and performs better than several supervised techniques. The proposed method, being lightweight, simple, and robust, offers the suitability of being adopted as a tool for online rumor identification.

**Keywords:** Unsupervised, Early detection, Clustering, Content features, Social features

## 1 Introduction

Social media facilitates the feasible propagation of information to aid communication, which has resulted in the proliferation of rumors on the internet. People experience a shortage of time in the fast-paced world, and social media serves as an instantaneous news dissemination platform. In addition to being the peoples' news channel, dependence on social media to obtain and convey information is beneficial for journalists to furnish the masses with exclusive breaking news readily. [1]–[3]. Twitter is one of such platforms widely employed as the breeding ground of information, outperforming several traditional sources in projecting news [4]. People's reliance on the news shifts from traditional to social news media, making them vulnerable to rumors and misinformation [5]. Users' gullibility towards online information has promoted rumors to spread wider and faster [6]. Tam [7] asserts that the decentralized architecture of social platforms is a cause allowing the spread of rumors without moderation. Cai et al. [8] define rumor to be a statement that is supposed to be false. DiFonzo and Bordia [9] state rumors to be unverified and instrumentally relevant information statements in circulation. Vosoughi et al. [10] simplify the definition as an unproven statement initiated from different sources that disseminate periodically from one point to another point in a network. Bai et al. [6] define rumor as unverified information at the time of its posting, causing anxiety and panic to varying degrees. The propagation of rumors has witnessed severe impacts worldwide. The hacking of a Twitter account in April 2013 emanated a rumor about two explosions at the White House injuring Barrack Obama, resulting in significant economic losses, including a $136.5 billion loss in the S&P 500 market [11]. Online rumors impact various spheres of life by defaming individuals and organizations, causing mental trauma, disrupting financial markets, and derailing day-to-day events.

Discerning true information amongst rumors is a complex task for the public [12]. Huge social media engagements and the dynamic nature of platforms make it challenging to assess the veracity of continuously pouring information. The most apparent information available through social media microblogs is the content information which has motivated many scholars to develop rumor detection models employing content-based features. The progress evolves through the use of TF-IDF models, machine learning, and deep learning models [13]. Content information consists of syntactic, lexical, and semantic features [6]. However, social media posts are short and semantically weak. Such models depending solely on content features provided unsatisfactory results, highlighting the need to examine structural features for the task. These structural features commonly referred to as context-based features, constitute user information [14] and network information [15].

We hypothesize that when tweet attributes are mapped onto an n-dimensional space, rumor and non-rumor posts can be grouped into two distinct clusters. This work employs either content or social features and their combination to perform unsupervised clustering. using Radial Basis Function (RBF) spectral clustering, Nearest Neighbour (NN) spectral clustering, K-means clustering, and Fuzzy C means clustering. We obtain promising results using the proposed approach to perform early segregation of rumor and non-rumor tweets. The accuracies obtained are ~20-25% higher than the existing state-of-the-art techniques.

The organization of the paper is as follows: we discuss the related literature in Section 2. Section 3 emphasizes the features employed and the proposed methodology with its mathematical background. Section 4 discusses the dataset used, results obtained, and the comparison with existing baseline methodologies. In section 5, we conclude by discussing future prospects.

## 2 Related Work

Knapp et al. (1994) initially classified rumors depending upon two perspectives: (a) impact perspective, which relates to the type of impact or effect a rumor has on the world, and (b) classification perspective where rumors are labeled with the intent of classification. Within the first category, Pipe-dream rumors lead to wishful thinking, bogy rumors cause anxiety or fear, and wedge-driving rumors lead to hatred generation. The second category defines apriori rumors as time-old and have been discussed for a long duration of time. In contrast, new emerging rumors are those arriving through breaking news events.

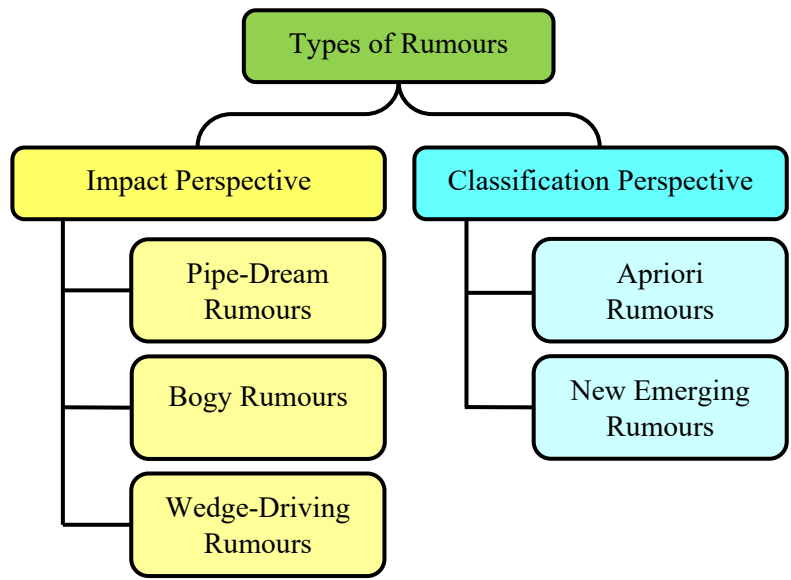

Figure 1: Rumour classification according to Knapp et al.

Mechanisms towards effective rumor identification and detection posit several alternatives ranging from the use of content features, contextual features, user-based features, network-based features, and psycholinguistic features. Various authors have listed out several baseline features employed for online rumor classification. Varshney and Vishwakarma [16] analyze a list of 81 handcrafted features for multimedia-based rumor classification. Figure 2 categorizes these features into five significant types depicting their sub-types.

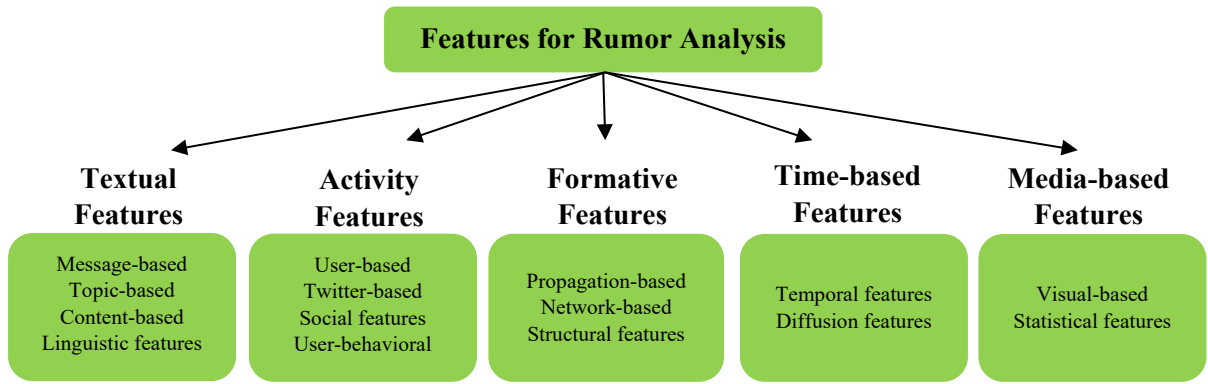

Figure 2: Feature classification for rumor analysis

Existing studies are employing machine learning for automatic classification of rumors, and non-rumors focusing on supervised learning. Binary classification remains the most-used mechanism that requires the training of a classifier [17]–[19]. Meel and Vishwakarma [20] survey the state-of-the-art in rumor propagation and detection by analyzing ongoing developments, challenges and future directions. Very few studies have examined the need for spontaneous rumor detection and proposed unsupervised methods. Studies towards automatic detection of online rumors are limited [21]. Existing methods are capable of distinguishing between rumors and non-rumors when the labels are known a priori. However, very few approaches can identify a post as a rumor as soon as it gets online.

Most of the existing works rely on supervised machine learning algorithms supported by feature extraction. Studies have broadly classified features as message-based, user-based, topic-based, and propagation-based [18]. Kwon et al. [22], [23] added to this list by introducing the use of linguistic, structural, and temporal features. Earlier works have collectively used a set of social features, including text, user information, rumor topic, and propagation-based features [18], following the individual methods by earlier works [24]–[26]. They experimented with a J48 decision tree classifier on a Twitter dataset [26], obtaining up to 80% precision and recall.

Yang et al. [28] emphasized that propagation features alone were insufficient in identifying rumors, whereas content features play a beneficial role. Also, noting that account features have critical participation, they employed a combination of content, account, and propagation features to classify rumor posts using an SVM with RBF Kernel, increasing accuracy by ~5%.

Zhao et al. [11] assumed that social media users should inquire about tweet veracity in case of rumors that promoted content-based features. Bhattacharjee et al. [13] followed an event-based approach by characterizing events under a particular topic. They classified rumors and non-rumors for each event separately. The introduction of structural features as contextual information to classification models further improved the detection performance [15].

Ma et al. [29] incorporated time-series modeling to utilize temporal features of rumor propagation on social media. The method focuses on the social context variation over time for social media messages by generating timestamps and using Dynamic Series-Time Structure (DSTS) with popular machine learning algorithms: Decision Tree, SVM, and Random Forest. The novel method captures salient time features demonstrating substantial improvement in supervised learning for rumor classification. Sampson et al. [30] also employed temporal features experimenting with Dynamic Tree (DT) time structure, DSTS, hybrid SVM, and RBF kernel, significantly leveraging the task of early classification of online rumors.

Table 1: Summary of related work

| Ref | Method | Algorithm | Dataset | Result |
|---|---|---|---|---|
| [28] | Machine Learning | SVM with RBF kernel | Sina Weibo | - |
| [29] | Machine Learning | Decision Tree, SVM, Random Forest | Castillo dataset | Accuracy: 86.7% |
| [31] | Machine Learning | Graph-based SVM with Random walk kernel and RBF kernel. | Sina Weibo | Accuracy: 91.3% |
| [21] | Machine Learning | SVM, Conditional Random Field, Naive Bayes, Random Forest | PHEME | Precision: 66.7 |
| [30] | Machine Learning | DT (Dynamic Tree Time Structure, DSTS (Dynamic Series Time Structure), Hybrid DSTS with SVM and RBF kernel | Castillo dataset, Snopes | Accuracy: 98.5% |
| [32] | Machine Learning | SVM, Logistic Regression, K-Star, and Random Forest. | MediaEval 2015 | Accuracy: 83.6% |
| [33] | Machine Learning | One Class Classification (OCC) | Zubaigaset, Kwonset | F1-score: 74%, 93% |
| [34] | Machine Learning | SVM, Decision Tree, Random Forest, K-Nearest Neighbour, Gradient Boost Decision Tree, Xgboost. | Sina Weibo | F1-score: 82.5% |
| [35] | Deep Learning | Dual convolutional neural networks (CNN) | Twitter | 5-35% improvement over baselines |
| [36] | Deep Learning | GRU, RNN, LSTM ensemble | PHEME | 7.9% improvement over baselines |

Wu et al. [31] introduced a novel graph-kernel-based approach with an SVM classifier to prioritize propagation and sematic features for the rumor classification task. Their algorithm uses a hybrid kernel function, a combination of radial basis function and a random walk graph kernel. The technique utilizes topic-based features, sentiment scores, and user profiles in addition to propagation features. They concluded that the repost patterns of rumors and non-rumors are very different on microblogging platforms, thus establishing it as an essential feature and making their algorithm suitable for early detection of rumors.

Zubiaga et al. [21] proposed a novel sequential classifier to leverage microblog context for rumor classification using Conditional Random Fields (CRF). Their technique improves performance over traditional machine learning models such as SVM, Naïve Bayes, and Random Forest.

Jin et al. [37] highlighted the importance of visual content in automation rumor classification. Their multimodal approach uses textual, visual, and statistical features with SVM, Logistic Regression, K Star, and Random Forest algorithms, outperforming only-text approaches.

Fard et al. [33] gave a new direction to rumor detection by developing a novel one-class classification (OCC) approach in which an algorithm is treated with items of only rumor class. The authors extract a set of 86 features from the rumor category of the Twitter dataset to train traditional machine learning algorithms. The method aims to remove the annotation bias caused due to manual annotation and volition of rumors and non-rumors.

Geng et al. [34] proposed a new set of features, namely, Words with Guidance (WG), Words with Menace (WM), Suspected Topic (ST), Recognition of Information (RI), Degree of Attention to Users (DAU), and Credit Rating (CR). They introduced SMOTE, a resampling algorithm to counter the impact of imbalanced datasets. The improved feature selection and resampling positively influenced the effectiveness of their rumor detection system.

Progressing further, researchers are employing deep neural networks for the rumor detection task owing to their efficacy and robustness. However, such an approach is not successful for the early detection of rumors because online rumor data is sparse and does not provide a considerable training dataset to the neural networks. Santhoshkumar & Babu [35] adapted a deep learning approach using a dual Convolutional Neural Network (CNN), which uses a certainty factor that relies on implicit features: temporal, content, and propagation features. Their proposed approach efficiently combines the classification power of neural networks with that of inherent features of microblogs.

Kotteti et al. [36] advanced in this direction by proposing a deep-learning-based ensembling architecture constituting of Gated Recurrent Unit (GRU), Recurrent Neural Network (RNN), and Long Short-Term Memory network (LSTM) depending upon time-series representations of Twitter data.

## 3  Proposed Methodology

We formulate the rumor detection task as an unsupervised classification problem, considering that the tweets are unlabelled, and the algorithm is solely responsible for differentiating rumors from non-rumors with no apriori knowledge. The experimental dataset contains the set of tweets, $T = \{t_1, t_2, t_3,…,t_n\}$, with each tweet labelled as rumour (R) or non-rumour (NR) and belonging to an individual event among the set of events $E = \{e_1, e_2, e_3,…,e_9\}$. The problem statement is defined as a task automatically assigning each tweet from the set T, a distinct cluster depending upon their characteristic features. Initially, the algorithm inputs tweets alone, without their labels, and it groups them into two separate clusters based on their similarity with their nearest tweets. Notably, the algorithm is responsible for rumor detection, i.e., distinguishing rumors from non-rumor tweets but not for veracity analysis (determining whether a rumor is true, false, or unverified), which is a separate task according to Zubiaga et al.'s rumor classification system [21]. We extract two sets of features: $F_c$ (content features) and $F_s$ (social features) for each tweet in T. Extracted features are normalized and undergo dimensionality reduction to reduce the feature set into two principal components. To perform unsupervised classification, the set of tweets from each event E is fed to a clustering algorithm to automatically separate rumors from non-rumors based on their differentiable characteristics (features). Figure 3 depicts the workflow of the proposed framework.

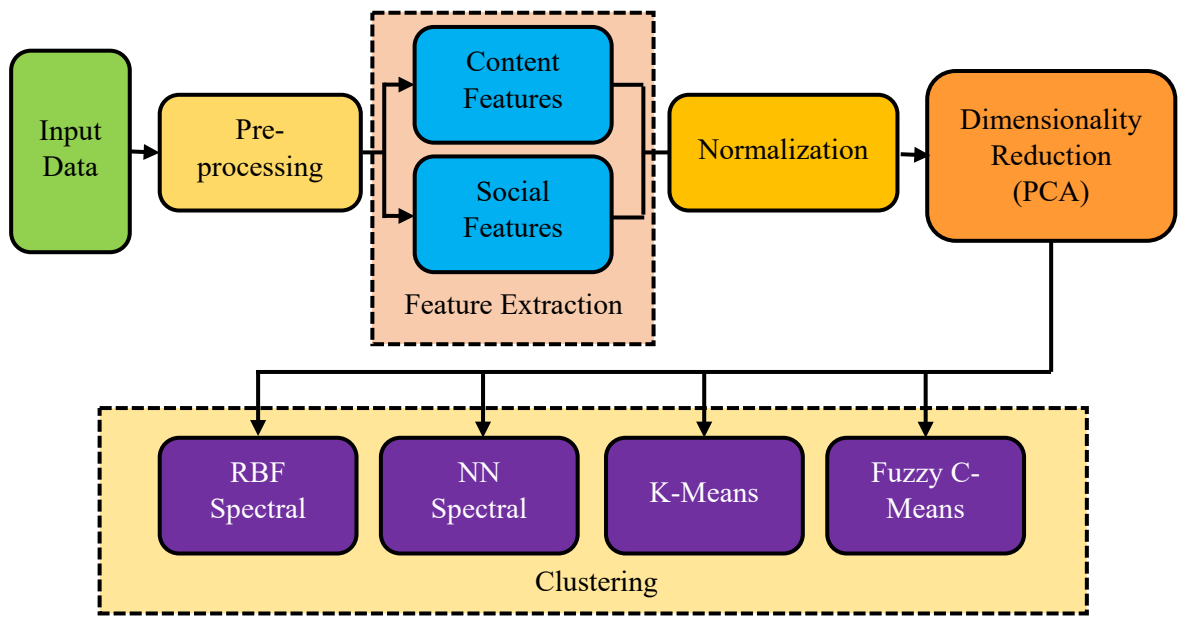

Figure 3: Proposed methodology for unsupervised rumor detection

**Feature Extraction:** Rumours and non-rumors demonstrate spotting differences in their characteristics. We use sixteen features in total, with five content-based features and eleven social features, as depicted in Table 2. Existing studies state that the content of rumor and non-rumor posts differ substantially in their writing styles. Specifically, such posts differ in their word count. It is observed that rumor posts constitute a greater number of words than non-rumors. Figure 4 illustrates the box plots with a 95% confidence interval to differentiate the value ranges of rumor and non-rumor tweets for each feature used. Additionally, users posting false information have been observed to use higher periods, question marks, capital letters, and hashtags. We use these remarkable characteristics as content-based features to distinguish rumors from non-rumor tweets. Social features play a significant role in distinguishing such news items. We use the following social features in the proposed framework: favorite count, retweet count, user's default profile image, verified user, possibly sensitive, time difference, followers count, listed count, status count, friends count, and user's favorite count. The box plot for the possibly_sensitive feature demonstrates that non-rumors contain more sensitive content than rumors. Also, there is a considerable difference between the time difference of rumor and non-rumor posts which is calculated by subtracting the tweet time from the account creation time.

Table 2: Description of content and social features used in the proposed methodology

| Feature | Description |
|---|---|
| *content features* | |
| Word Count | Number of words in a tweet/post |
| Period Present | Parameter to verify if periods (.) are present in a tweet |
| Question Mark | Parameter to verify if question marks (?) are present in a tweet |
| Capital Ratio | The ratio of capital letters to small letters in a tweet |
| Hashtag Ratio | Parameter to verify if hashtags (#) are present in a tweet |
| *social features* | |
| Time Difference | Difference between the time a user account is created and the post is shared |
| Possibly Sensitive | Parameter to indicate if a tweet contains sensitive content |
| Favorite Count | Parameter to indicate the count that users like a tweet |
| Retweet Count | The number of times a tweet is retweeted |
| Default Profile Image | Parameter indicating if a user has uploaded a profile picture or not |
| Verified | Parameter to indicate if a user has a verified Twitter account or not |
| Follower Count | The number of followers of a Twitter user account |
| Listed Count | The number of public lists a Twitter user is a member of |
| Status Count | The total number of tweets of a Twitter user since its account creation |
| Friend Count | The total number of Twitter users an account follows |
| Total Favorite Count | The total number of likes on the tweets of a Twitter account |

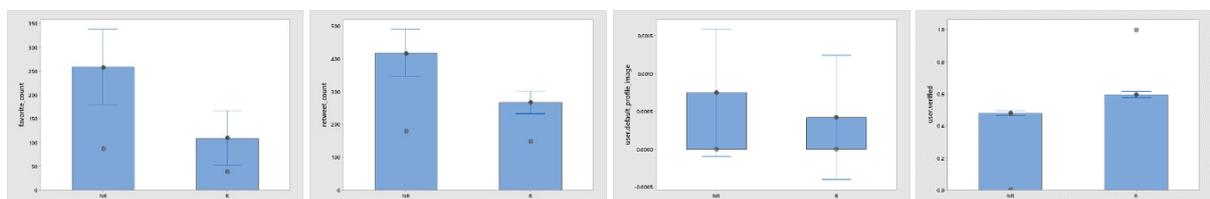

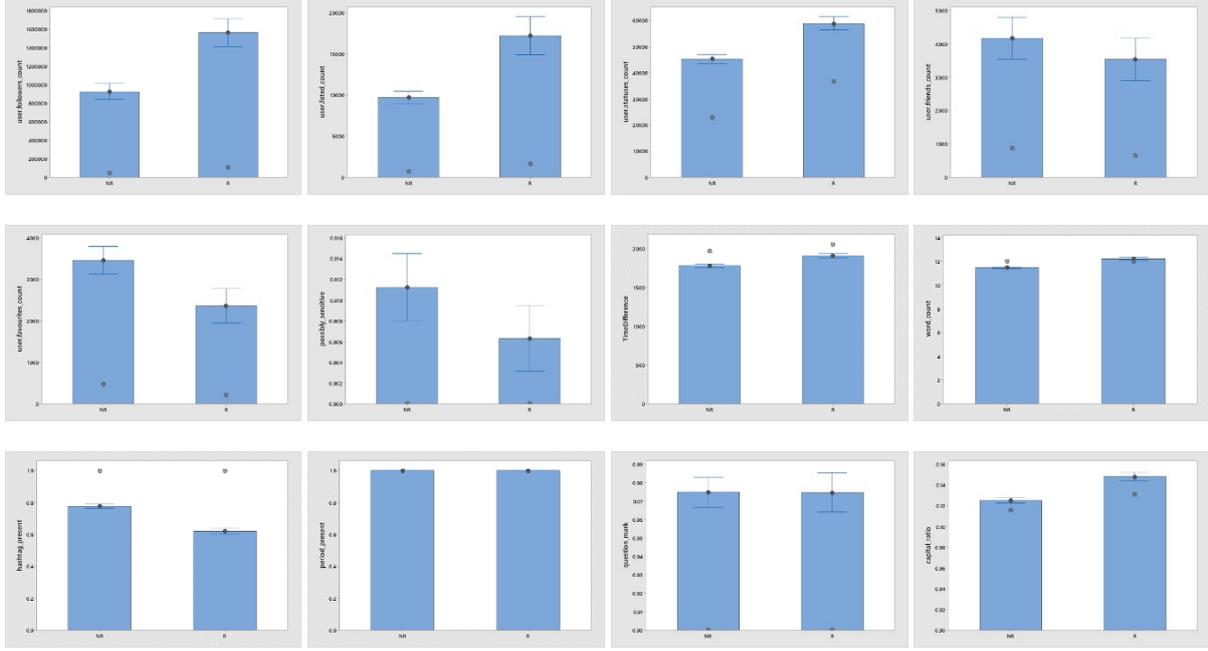

Figure 4: Box plots for contrasting characteristics between attributes features of rumours and non-rumours

**Normalization:** Features in the extracted data lie within different scale ranges, which weakens the model's efficacy. If different features having discrete values lie on varied scales, the importance of lower-scale attributes is undermined compared to larger-scale attributes, thereby reducing the importance of low-scale features. It is essential to bring the values of multiple features on the same scale within a fixed range with the same minimum and maximum limits. We scale the attribute values within a scale of 0 to 1 using Min-Max normalization. A linear transformation is performed on the actual data according to Eq. 1, where A is the attribute data. Min(A) and Max(A) denote the attributes' minimum and maximum absolute values, respectively. The original values and normalized values are represented by v and v', respectively. Here, new_min(A) and new_max(A) are used to set the minimum and maximum values of the new scale, which in our case are 0 and 1, respectively.

$$v' = \frac{v - \min(A)}{\max(A) - \min(A)} (new\_\max(A) - new\_\min(A)) + new\_\min(A) \quad (1)$$

**Dimensionality Reduction:** Large feature set leads to an increased dataset dimension, which decreases a model's efficiency and requires more data generalization. We perform dimensionality reduction using Principal Component Analysis (PCA) for data's complexity reduction. PCA is an unsupervised algorithm that transforms high-dimensional data into a low-scale subspace, retaining relevant information. It works on the principle of finding correlations among attributes to identify functional patterns. The algorithm looks for directions with maximum variance and transforms them as the orthogonal axes of the new subspace. We reduce a feature set of sixteen attributes with $D = 16$ to a subspace with two principal components ($d = 2$). The PCA dimensionality reduction algorithm is given below:

---
**Algorithm 1** Dimensionality reduction with PCA
1. **Input:** a D-dimensional feature set $X = \{x_1, x_2, \ldots, x_N\}$ and the new target dimensionality, d, of lower subspace, where $d \leq D$
2. **Calculate** the mean $\bar{x} = \frac{1}{N}\sum_{i=1}^{N} x_i$
3. **Find** the covariance matrix given by $Cov(x) = \frac{1}{N}\sum_{i=1}^{N}(x_i - \bar{x}_i)(x_i - \bar{x}_i)^T$
4. **Decompose** $Cov(x)$ into its eigenvectors $\varepsilon_1, \varepsilon_2, \ldots, \varepsilon_D$ and eigenvalues $\lambda_1, \lambda_2, \ldots, \lambda_D$, sorting the eigenvalues in descending order.
5. **Calculate** the new lower dimension, y, as:
$$y = (\varepsilon_1^T(x - \bar{x}), \varepsilon_2^T(x - \bar{x}), \ldots, \varepsilon_d^T(x - \bar{x}))^T$$
where $x \in \mathbb{R}^D$ and $y \in \mathbb{R}^d$.

---

**Unsupervised Clustering:** An unsupervised clustering algorithm is used to obtain two principal components from the set of features. In our approach, we experiment with four types of clustering: RBF Spectral clustering, NN Spectral clustering, K-means clustering, Fuzzy C means clustering.

*Spectral Clustering:* Spectral clustering is an advancing algorithm that performs better than several traditional clustering algorithms. We consider the data points (tweets) to be nodes of a graph, and the problem formulates to be a graph partitioning situation. We group data points by computing the connectivity between them, irrespective of the distance. An Affinity Matrix is created to determine the similarity between two points within the dataset. A Laplacian vector transforms the data points into a two-dimensional plot that can be easily segregated into two separate clusters. We experiment with two types of spectral clustering: Radial Basis Function (RBF) spectral and Nearest Neighbour (NN) spectral clustering, both differing on the principle of their affinity measures. The pseudocode for the algorithm is as below:

---
**Algorithm 1** Spectral Clustering
1. Input the set of data points, $D = \{d_1, d_2, \ldots, d_n\}$
2. Create the Affinity Matrix $A_{ij}$ as below:
$$A_{ij} = e^{\frac{-\|s_i - s_j\|^2}{2\sigma^2}}$$
Where $i \neq j$, and $A_{ij} = 0$.
3. Define a diagonal matrix:
$$D_{ii} = \sum_K a_{ik}$$
4. Create a matrix $L$ as given below:
$$L = D^{-1/2} A D^{-1/2}$$
5. Stack k largest eigenvectors of L to form the columns of the new matrix $X$:

$$X = \begin{bmatrix} | & | & & | \\ x_1 & x_2 & \ldots & x_k \\ | & | & & | \end{bmatrix}$$

6. Renormalize each of $X's$ rows to have a unit length and get a new matrix $Y$. Cluster rows of $Y$ as points in $R^k$

*K-Means Clustering:* K-means clustering derives K distinct clusters from the feature set where each data point belongs to a single cluster. It aims to maintain the homogeneity within each cluster by keeping the most similar points together. The number of clusters to be derived has to be fixed apriori. For the given set of data points, we assign $K = 2$, and the algorithms partitions the observations into two clusters C1 and C2. The aim is to designate two centers, one for each cluster, such that they are at the farthest distance from each other. The objective is to minimize the variance among the observations within a cluster by picking each data point and setting it near the closest center. After all the data points have been assigned to the nearest center, two new centroids are recalculated, and the entire steps are iterated. The two centers gradually shift their location with each iteration until they fix their positions as the final centroids of the clusters.

**Algorithm 1** K Means Clustering
1. Randomly select two centroids, one for each cluster, C1 and C2, such that $\mu_1, \mu_2 \in \mathbb{R}^n$.
2. Iterate until convergence: {
   For every i, set
   $$c^i = arg \min_j \|x^i - \mu_j\|^2.$$
   For every j, set
   $$\mu_j = \frac{\sum_{i=1}^m 1\{c^i = j\}x^i}{\sum_{i=1}^m 1\{c^i = j\}}$$
   }

*Fuzzy C-Means Clustering:* Fuzzy C-Means clustering iteratively works on fuzzy logic where we assign a probabilistic score between 0 and 1 to each data point, determining the possibility of a data point belonging to a cluster. It aims to provide the strength of the relationship between the point and the cluster based on the distance from the cluster's centroid. The sum probability score of all data points related to a cluster is equal to one.

**Algorithm 1** Fuzzy C Means Clustering
3. For i=2, initialize the membership matrix, given by:
$$\sum_{j=1}^C \mu_j(x_i) = 1$$
4. Find the centroid $C_j$ of the cluster given by:

$$C_j = \frac{\sum_i [\mu_j(x_i)]^m x_i}{\sum_i [\mu_j(x_i)]^m}$$

Where $m$ is the fuzzification parameter, $m \in [1.25, 2]$.

5. Calculate the Euclidean distance between each data point and the centroid $C_j$ as follows:

$$D_i = \sqrt{(x_2 - x_1)^2 + (y_2 - y_1)^2}$$

6. Recalculate the updated membership matrix, $\mu_j(x_i)$:

$$\mu_j(x_i) = \frac{[\frac{1}{d_{ji}}]^{1/m-1}}{\sum_{k=1}^{C} [\frac{1}{d_{ki}}]^{1/m-1}}$$

7. Iterate through steps 2, 3, and until centroids are fixed.

## 4 Experimental Result Analysis

### 4.1 Dataset

In this paper, we use the only available large-scale rumor classification dataset, PHEME [21]. Table 3 depicts the description of the PHEME dataset. The dataset consists of new-emerging news on Twitter from nine breaking-news events that are unknown apriori, consisting of rumor and non-rumor classes. A count of 6425 tweets was scraped from the Twitter API, with 2402 tweets as rumors and 4023 tweets annotated as non-rumors. The breaking news events in the dataset are described as follows:

***Charlie Hebdo:*** Two brothers forced their way into the offices of the French satirical weekly newspaper Charlie Hebdo, killing 11 people and wounding 11 more, on January 7, 2015.

***Ebola Essien:*** A Twitter user posted on October 12, 2014, stating that the AC Milan footballer Michael Essien had contracted Ebola. The report was later denied by the footballer and thus exposed as a hoax.

***Ferguson:*** Citizens of Ferguson in Missouri, USA, protested after the fatal shooting of an 18-year-old African American, Michael Brown, by a white police officer on August 9, 2014.

***Germanwings Crash:*** A passenger plane from Barcelona to Dsseldorf crashed in the French Alps on March 24, 2015, killing all passengers and crew. The plane was ultimately found to have been deliberately crashed by the co-pilot.

***Gurlitt:*** A rumor in November 2014 that the Bern Museum of Fine Arts was going to accept a collection of modernist masterpieces kept by the son of a Nazi-era art dealer. The museum did end up accepting the collection, confirming the rumors.

***Ottawa Shooting:*** Shootings occurred on Ottawa's Parliament Hill, resulting in the death of a Canadian soldier on October 22, 2014.

***Prince Toronto:*** A rumor started circulating on November 3, 2014, that the singer Prince would play a secret show in Toronto that night. Some people even queued at the venue to attend the concert, but the rumor was later proven false.

***Putin Missing***: Numerous rumors emerged in March 2015 when the Russian president Vladimir Putin did not appear in public for ten days. He spoke on the 11th day, denying all rumors that he had been ill or was dead.

***Sydney Siege:*** A gunman held hostage ten customers and eight employees of a Lindt chocolate cafe located at Martin Place in Sydney on December 15, 2014.

Table 3: Number of items in each event in PHEME dataset

| Events | Rumours | Non-rumours | Total |
|---|---|---|---|
| Charlie Hebdo | 458 | 1621 | 2079 |
| Ebola Essien | 14 | 0 | 14 |
| Ferguson | 284 | 859 | 1143 |
| Germanwings Crash | 238 | 231 | 469 |
| Gurlitt | 61 | 77 | 138 |
| Ottawa Shooting | 470 | 420 | 890 |
| Prince Toronto | 229 | 4 | 233 |
| Putin Missing | 126 | 112 | 238 |
| Sydney Siege | 522 | 699 | 1221 |
| Total | 2402 | 4023 | 6425 |

## 4.2 Evaluation Metrics

To evaluate the performance of our approach, we use the confusion matrix to calculate four essential parameters: accuracy, precision, recall, and F1-scores, defined in Eq. 2-5. According to the confusion matrix, True Positive (TP) and False Negative (FN) are the news items correctly classified, whereas True Negative (TN) and False Positive (FP) denote wrongly clustered items.

|  | Rumour | Non-Rumour |
|---|---|---|
| Rumour | True Positive (TP) | True Negative (TN) |
| Non-Rumour | False Positive (FP) | False Negative (FN) |

$$Accuracy = \frac{TP+TN}{P+N} \qquad (2)$$

$$Precision = \frac{TP}{TP+FP} \qquad (3)$$

$$Recall = \frac{TP}{TP+FN} \qquad (4)$$

$$F1-score = \frac{2\,TP}{2\,TP+FP+FN} \tag{5}$$

## 4.3 Results

Results achieved on the nine events in the PHEME dataset are listed in Tables 4-12. Each table shows the results achieved from each clustering algorithm using content features, social features, and a combination of both. For the Charlie Hebdo event, we observe that RBF clustering gave the highest results with 77.97% accuracy using social features ($F_S$) and a combination of content and social features ($F_{C+S}$). Whereas solely content features achieve 67.05% accuracy using RBF spectral clustering. Significantly high results are observed using all four algorithms with $F_{C+S}$. However, K-means displays the highest recall of 78.94% using $F_S$ and $F_{C+S}$. Clearly, the combination of social and content features is superior to using any of these features alone.

The Ebola Essien event is a relatively small and imbalanced dataset with only 14 rumors and no non-rumor tweets. This makes it challenging to analyze the performance of algorithms. In such a case, the precision cannot be calculated as data points from one of the classes are missing. Also, it makes the recall and F1-score to be zero. The comparison can be made based on accuracy scores only, which is the highest, again, using RBF spectral clustering with $F_{C+S}$ giving 85.71%. NN clustering, K-means, and Fuzzy C means achieve 78.57 accuracies when experimented with $F_{C+S}$. Also, using only social features, K-means and Fuzzy C means observe 78.57% accuracy. Content features have obtained a substantially lower result of 28.57% with all four algorithms.

For the Ferguson event, the highest accuracy is 75.15% with 99.77% precision and 85.79% F1-score using RBF clustering with only social features. The amalgamation of content and social features obtains marginally lower results of 74.45% accuracy, 97.56% precision, 75.56% recall, and 85.16% F1-score using K-means and Fuzzy C means algorithms. On the other hand, solely content features score substantially low with only 57.39% accuracy using RBF spectral clustering.

Observing the scores achieved for Germanwings crash event, the highest accuracy of 99.36% is demonstrated by Fuzzy C means algorithm using only content features with a precision of 100%, 98.72% recall, and 99.35% F1-score. The second highest results show a vast difference by only 52.03% accuracy using social features with NN spectral clustering. The rest of the results are close to 45-50% accuracy, which demonstrated Fuzzy C means as the best clustering algorithm for this event. In contrast to other events, content features have shown a significant improvement in performance.

Results for the Gurlitt event also strengthen the efficacy of Fuzzy C means using content features only. We observe an accuracy of 94.2%, 89.62% precision, 100% recall, and 94.52% F1 score. Contrastingly, $F_S$ and $F_{C+S}$ receive ~58% accuracies which are comparatively lower than Fuzzy C means.

The Ottawa shooting event also holds the highest results using Fuzzy C means with content features. We observed 97.19% accuracy, a 100% precision with 94.38% recall, and a 97.11% F1 score. The second highest accuracy is 50.11%, observed by K means using $F_S$ and $F_{C+S}$.

We obtained mixed results for the Prince Toronto event. NN spectral clustering demonstrates 57.51% accuracy using $F_{C+S}$ and 56.65% accuracy using $F_S$. However, the precision, recall, and F1 scores are very low. RBF spectral and K means demonstrate 52.36% accuracy using only content features. The precision in most cases receives a 100% score, whereas the recall is nearly 2% which highlights the imbalance in the dataset.

On the Putin missing dataset, the highest performance is displayed by NN spectral clustering using social features. The precision and F1 scores for K-means and Fuzzy C means were 99.11% and 64.16% for both $F_S$ and $F_{C+S}$. The combined set of features suggest a comparable performance with 62.61% accuracy using NN spectral clustering.

The highest accuracy score for the Sydney siege event is 58.31% by Fuzzy C means with social features. Performance using $F_C$ and $F_{C+S}$ are also similar with 56.92% and 57.25% accuracies. RBF spectral is clearly the best performing algorithm in this case.

Analyzing the overall performance, we see that the performances of social features and the combination of content and social features are comparably similar. However, content features show substantial improvement with the Fuzzy C means clustering algorithm. In many cases, K-means and Fuzzy C means perform equally well. RBF spectral clustering algorithm, however, maintains consistency throughout all events by delivering high results.

Table 4: Results on Charlie Hebdo event using four algorithms with content and/or social features

| Clustering | Content | | | | Social | | | | Content + Social | | | |
|---|---|---|---|---|---|---|---|---|---|---|---|---|
| | A | P | R | F1 | A | P | R | F1 | A | P | R | F1 |
| RBF Spectral | **67.05** | **84.08** | 76.15 | **79.92** | **77.97** | **100** | 77.9 | **87.62** | 77.97 | 100 | 77.97 | **87.62** |
| NN Spectral | 63.44 | 73.6 | **78.23** | 75.84 | **77.97** | 98.21 | 78.77 | 87.42 | 76.48 | 97.35 | 77.96 | 86.58 |
| K-Means | 48.44 | 51.33 | 74.62 | 60.82 | 76.77 | 95.74 | **78.94** | 86.53 | 76.77 | 95.74 | **78.94** | 86.53 |
| Fuzzy C | 27.95 | 35.84 | 55.92 | 43.68 | 23.23 | 4.26 | 61.06 | 7.96 | 76.77 | 95.74 | **78.94** | 86.53 |

Table 5: Results on Ebola Essien event using four algorithms with content and/or social features

| Clustering | Content | | | | Social | | | | Content + Social | | | |
|---|---|---|---|---|---|---|---|---|---|---|---|---|
| | A | P | R | F1 | A | P | R | F1 | A | P | R | F1 |
| RBF Spectral | **28.57** | - | 0 | 0 | 14.29 | - | 0 | 0 | **85.71** | - | 0 | 0 |
| NN Spectral | **28.57** | - | 0 | 0 | 21.43 | - | 0 | 0 | 78.57 | - | 0 | 0 |
| K-Means | **28.57** | - | 0 | 0 | **78.57** | - | 0 | 0 | 78.57 | - | 0 | 0 |
| Fuzzy C | **28.57** | - | 0 | 0 | **78.57** | - | 0 | 0 | 78.57 | - | 0 | 0 |

Table 6: Results on Ferguson event using four algorithms with content and/or social features

| Clustering | Content | | | | Social | | | | Content + Social | | | |
|---|---|---|---|---|---|---|---|---|---|---|---|---|
| | A | P | R | F1 | A | P | R | F1 | A | P | R | F1 |
| RBF Spectral | **57.39** | **73.57** | 70.85 | **72.19** | 75.15 | **99.77** | 75.24 | **85.79** | 24.67 | 0 | 0 | 0 |
| NN Spectral | 50.83 | 53.67 | **73.76** | 62.13 | 67.19 | 79.98 | **77.19** | 78.56 | 32.81 | 20.02 | 67.98 | 30.94 |
| K-Means | 49.17 | 55.88 | 70.38 | 62.3 | 74.45 | 97.56 | 75.56 | 85.16 | **74.45** | **97.56** | **75.56** | **85.16** |
| Fuzzy C | 25.11 | 33.41 | 50.26 | 40.14 | 25.55 | 2.44 | 61.76 | 4.7 | **74.45** | **97.56** | **75.56** | **85.16** |

Table 7: Results on Germanwings Crash event using four algorithms with content and/or social features

| Clustering | Content | | | | Social | | | | Content + Social | | | |
|---|---|---|---|---|---|---|---|---|---|---|---|---|
| | A | P | R | F1 | A | P | R | F1 | A | P | R | F1 |
| RBF Spectral | 45.42 | 77.92 | 46.75 | 58.44 | 49.47 | **100** | 49.36 | **66.09** | 49.47 | 100 | 49.36 | 66.09 |
| NN Spectral | 44.14 | 71.86 | 45.73 | 55.89 | **52.03** | 13.42 | **55.36** | 21.6 | 48.4 | 84.42 | 48.63 | 61.71 |
| K-Means | 38.81 | 45.45 | 39.47 | 42.25 | 47.76 | 93.51 | 48.43 | 63.81 | 47.76 | 93.51 | 48.43 | 63.81 |
| Fuzzy C | **99.36** | **100** | **98.72** | **99.35** | 47.76 | 93.51 | 48.43 | 63.81 | 47.76 | 93.51 | 48.43 | 63.81 |

Table 8: Results on Gurlitt event using four algorithms with content and/or social features

| Clustering | Content | | | | Social | | | | Content + Social | | | |
|---|---|---|---|---|---|---|---|---|---|---|---|---|
| | A | P | R | F1 | A | P | R | F1 | A | P | R | F1 |
| RBF Spectral | 60.14 | 81.82 | 60.58 | 69.61 | 57.25 | **100** | 56.62 | 72.3 | 58.7 | 93.51 | **58.06** | 71.64 |
| NN Spectral | 39.13 | 48.05 | 45.68 | 46.84 | 54.35 | 42.86 | 63.46 | 51.16 | 45.65 | 58.44 | 51.14 | 54.55 |
| K-Means | 37.68 | 22.08 | 39.53 | 28.33 | **57.97** | **100** | **57.04** | **72.64** | 57.97 | **100** | 57.04 | **72.64** |
| Fuzzy C | **94.2** | **89.61** | **100** | **94.52** | 42.03 | 0 | 0 | 0 | 42.03 | 0 | 0 | 0 |

Table 9: Results on Ottawa Shooting event using four algorithms with content and/or social features

| Clustering | Content | | | | Social | | | | Content + Social | | | |
|---|---|---|---|---|---|---|---|---|---|---|---|---|
| | A | P | R | F1 | A | P | R | F1 | A | P | R | F1 |
| RBF Spectral | 61.8 | 33.57 | 69.8 | 45.34 | 47.42 | **100** | 47.3 | 64.22 | 47.42 | **100** | 47.3 | 64.22 |
| NN Spectral | 54.27 | 48.81 | 51.64 | 50.18 | 48.99 | 96.19 | 47.98 | 64.03 | 48.65 | 99.52 | 47.88 | 64.66 |
| K-Means | 41.12 | 45 | 39.21 | 41.91 | **50.11** | 97.62 | **48.58** | **64.87** | 50.11 | 97.62 | **48.58** | **64.87** |
| Fuzzy C | **97.19** | **100** | **94.38** | **97.11** | 49.78 | 96.67 | 48.39 | 64.5 | 49.78 | 96.67 | 48.39 | 64.5 |

Table 10: Results on Prince Toronto event using four algorithms with content and/or social features

| Clustering | Content | | | | Social | | | | Content + Social | | | |
|---|---|---|---|---|---|---|---|---|---|---|---|---|
| | A | P | R | F1 | A | P | R | F1 | A | P | R | F1 |
| RBF Spectral | **52.36** | **100** | **3.48** | **6.72** | 1.72 | 100 | 1.72 | 3.38 | 1.72 | 100 | 1.72 | 3.38 |
| NN Spectral | 39.91 | 50 | 1.43 | 2.78 | **56.65** | 25 | 1.01 | 1.94 | **57.51** | 25 | 1.03 | 1.98 |
| K-Means | **52.36** | **100** | **3.48** | **6.72** | 2.58 | **100** | 1.73 | 3.4 | 2.58 | **100** | **1.73** | **3.4** |
| Fuzzy C | 51.5 | **100** | 3.42 | 6.61 | 21.43 | **100** | **15.38** | **26.67** | 2.58 | **100** | **1.73** | **3.4** |

Table 11: Results on Putin Missing event using four algorithms with content and/or social features

| Clustering | Content | | | | Social | | | | Content + Social | | | |
|---|---|---|---|---|---|---|---|---|---|---|---|---|
| | A | P | R | F1 | A | P | R | F1 | A | P | R | F1 |
| RBF Spectral | 44.12 | **63.39** | 43.56 | **51.64** | 46.22 | 98.21 | 46.61 | 63.22 | 46.22 | 98.21 | 46.61 | 63.22 |
| NN Spectral | **49.16** | 57.14 | **46.72** | 51.41 | **65.97** | 56.25 | 66.32 | 60.87 | **62.61** | 60.71 | 60.18 | 60.44 |
| K-Means | 45.38 | 53.57 | 43.48 | 48 | 47.9 | **99.11** | 47.44 | **64.16** | 47.9 | **99.11** | **47.44** | **64.16** |
| Fuzzy C | 2.94 | 0 | 0 | 0 | 47.9 | **99.11** | 47.44 | **64.16** | 47.9 | **99.11** | **47.44** | **64.16** |

Table 12: Results on Sydney Siege event using four algorithms with content and/or social features

| Clustering | Content | | | | Social | | | | Content + Social | | | |
|---|---|---|---|---|---|---|---|---|---|---|---|---|
| | A | P | R | F1 | A | P | R | F1 | A | P | R | F1 |
| RBF Spectral | **56.92** | **99.28** | 57.12 | **72.52** | 57.25 | **100** | 57.25 | **72.81** | **57.25** | **100** | 57.25 | **72.81** |
| NN Spectral | 55.45 | 71.67 | **59.15** | 64.81 | 57.25 | 99 | 57.33 | 72.61 | **57.25** | 99 | **57.33** | 72.61 |
| K-Means | 49.55 | 54.22 | 56.15 | 55.17 | 41.69 | 3.58 | 39.68 | 6.56 | 41.69 | 3.58 | 39.68 | 6.56 |
| Fuzzy C | 7.29 | 12.73 | 14.57 | 13.59 | **58.31** | 96.42 | **58.2** | 72.59 | 41.69 | 3.58 | 39.68 | 6.56 |

## 4.4 Baseline Comparison

We compare our unsupervised approach with existing rumor detection mechanisms. Tables 13-15 compare results with the existing baselines using precision, recall, and F1 scores. The results of baselines are implemented on either content or social features or a combination of both. The highest results obtained are listed in the comparison tables.

**SVM:** Support vector machine using content and social features implemented with a cost coefficient derived by performing nested cross-validation.
**Random Forest:** An ensemble of multiple decision trees that use content features.
**Max Entropy:** A probabilistic classifier selects one classification model with the maximum entropy to fit the training data. It does not follow the naïve bayes principle of conditional independence of events.
**CRF:** Conditional Random Fields (CRF) is a sequential classifier that treats individual posts as a continuous thread of items. The tweets are modeled as a sequence or graph of rumors and non-rumors. [21] employ this technique to perform rumor detection on the PHEME dataset.
**CRF + HP:** Lathiya et al. [38] introduce a novel rumor detection technique using CRF with homophily features to measure the user engagement in rumor posts. Homophily features are derived by analyzing a user's network connections to see if they behave similarly to their connections in sharing rumor or non-rumor posts.

**CRF + RR:** Another technique by Lathiya et al. [38] that uses Rumour Ratio (RR) as the feature with CRF classifier. The ratio is calculated by dividing the total number of rumors spread by a Twitter account by the total number of posts spread by that account. It is given by Eq. 6

$$Rumour\ Ratio = \frac{number\ of\ rumours}{number\ of\ rumours + number\ of\ non-rumours} \qquad (6)$$

**TextCNN:** It is a one-dimensional text-based convolutional neural network used by Bai et al. [6].

**[38]:** Liu et al. [39] proposed a dynamic model for early real-time rumor detection by integrating a wide range of syntactical, semantic, stance-based, message-based, user-based, and indicator features. A random forest classifier is used as the real-time classification algorithm.

Tables 13-15 indicate the efficient performance of the proposed unsupervised approach over several existing state-of-the-art supervised rumor detection mechanisms. Results establish that the proposed supervised method performs rumor detection with significant accuracies. Since most of the datasets in PHEME are imbalanced with an unequal number of rumors and non-rumors, we use F1-score as a comparison measure because it disregards the dataset imbalance. For the Charlie Hebdo dataset, our maximum F1-score is 87.62% which is 11% higher than the maximum F1-score achieved by baseline TextCNN. We obtain an 85.79% F1-score for Ferguson, whereas TextCNN receives 70.2%, making our result ~15% higher. In the Germanwings crash dataset, we achieve the maximum F1-score of 99.35%, approximately 28% higher than the maximum baseline, CRF. In the Ottawa Shooting dataset, our approach yields a 97.11% F1-score surpassing the highest baseline of 72.3% (CRF+HP) by approximately 24%. In the Sydney Siege experiments, we received a maximum score of 72.81%, ~9% higher than the score obtained by CRF+HP (63.2%). Due to the absence of non-rumors in the Ebola Essien dataset, we compare the accuracies obtained on the dataset, and the highest accuracy of 85.71% is ~9% higher than the CRF+RR baseline. We obtain a 94.52% F1-score for the Gurlitt dataset, which is 23% greater than the highest baseline, TextCNN. For the Prince Toronto dataset, we receive a 26.67% F1-score making it 15% better than the TextCNN approach. We observe the highest F1-score of 64.16% for the Putin Missing dataset, 4% higher than the CRF+RR approach. All the above results demonstrate that the proposed framework certainly performs better than existing supervised learning algorithms. This eliminates the need to train complex algorithms with massive datasets; instead, rumor detection can be performed at early stages as and when they are received by grouping them into clusters based on their distinguishable characteristics.

Table 13: Baseline comparison for Charlie Hebdo, Ferguson, and Germanwings Crash events

| Classifier | Charlie Hebdo | | | Ferguson | | | Germanwings Crash | | |
|---|---|---|---|---|---|---|---|---|---|
| | P | R | F1 | P | R | F1 | P | R | F1 |
| SVM | 23.9 | 54.6 | 33.2 | 24.0 | 45.1 | 31.3 | 46.3 | 50.4 | 48.3 |
| RF | 21.5 | 20.3 | 20.9 | 25.4 | 12.7 | 16.9 | 43.8 | 2.9 | 5.5 |
| MaxEnt | 23.9 | 53.5 | 33.0 | 24.5 | 37.0 | 29.5 | 47.5 | 44.1 | 45.8 |
| CRF | 54.5 | 76.2 | 63.6 | 56.6 | 39.4 | 46.5 | 74.3 | 66.8 | 70.4 |
| CRF+HP | 48.8 | **80.6** | 60.8 | 55.4 | 39.8 | 46.3 | 68.6 | 75.2 | 71.7 |
| CRF+RR | 52.2 | 76.2 | 63.6 | 56.5 | 38.0 | 45.5 | 69.4 | 70.6 | 70.0 |
| TextCNN | 78.2 | 74.7 | 76.4 | 77.1 | 64.4 | 70.2 | 66.7 | 56.7 | 61.3 |
| [39] | 66.7 | 66.7 | 66.7 | 0 | 0 | 0 | 50.0 | **100** | 66.7 |
| Ours (C) | 84.08 | 76.15 | **79.92** | 73.57 | 70.85 | 72.19 | 100 | 98.72 | **99.35** |
| Ours (S) | **100** | 77.9 | 87.62 | **99.77** | 75.24 | **85.79** | 100 | 49.36 | 66.09 |
| C+S | **100** | 77.97 | 87.62 | 97.56 | 75.56 | 85.16 | 100 | 49.36 | 66.09 |

Table 14: Baseline comparison for Ottawa Shooting, Sydney Siege, and Ebola Essien events

| Classifier | Ottawa Shooting | | | Sydney Siege | | | Ebola Essien | | |
|---|---|---|---|---|---|---|---|---|---|
| | P | R | F1 | P | R | F1 | A | R | F1 |
| SVM | 49.6 | 42.8 | 45.9 | 43.5 | 48.5 | 45.8 | 37.4 | 0 | 0 |
| RF | 55.6 | 5.3 | 9.7 | 46.6 | 6.5 | 11.4 | 40.8 | 0 | 0 |
| MaxEnt | 51.2 | 40.9 | 45.4 | 42.5 | 42.9 | 42.7 | 32.46 | 0 | 0 |
| CRF | 84.1 | 58.5 | 69 | 76.4 | 38.5 | 51.2 | 75.8 | 0 | 0 |
| CRF+HP | 81.9 | 64.7 | 72.3 | 74.4 | 55.0 | 63.2 | 72.12 | 0 | 0 |
| CRF+RR | 83.5 | 59.4 | 69.4 | 75.4 | 51.0 | 60.8 | 76.82 | 0 | 0 |
| TextCNN | 76.2 | 52.5 | 62.2 | 74.7 | 47.3 | 57.9 | 57.12 | 0 | 0 |
| [39] | 90.9 | 50.0 | 64.5 | 64.7 | 33.3 | 44.0 | 39.42 | 0 | 0 |
| Ours (C) | **100** | **94.38** | **97.11** | 99.28 | 57.12 | 72.52 | 28.67 | 0 | 0 |
| Ours (S) | 97.62 | 48.58 | 64.87 | **100** | 57.25 | **72.81** | 78.57 | 0 | 0 |
| C+S | 97.62 | 48.58 | 64.87 | **100** | 57.25 | **72.81** | **85.71** | 0 | 0 |

Table 15: Baseline comparison for Gurlitt, Prince Toronto, and Putin Missing events

| Classifier | Gurlitt | | | Prince Toronto | | | Putin Missing | | |
|---|---|---|---|---|---|---|---|---|---|
| | P | R | F1 | P | R | F1 | P | R | F1 |
| SVM | 29.7 | 34.77 | 36.01 | 67.5 | 1.17 | 3.76 | 36.43 | 49.54 | 43.33 |
| RF | 25.51 | 18.74 | 21.28 | 74.12 | 3.41 | 5.93 | 40.61 | 37.15 | 37.21 |
| MaxEnt | 28.14 | 44.31 | 40.92 | 69.59 | 2.79 | 3.98 | 39.16 | 31.18 | 33.41 |
| CRF | 58.71 | 69.45 | 67.65 | 68.44 | 5.67 | 8.75 | 55.73 | 42.16 | 54.25 |
| CRF+HP | 52.9 | 74.12 | 69.14 | 75.71 | 6.71 | 9.82 | 59.27 | 54.16 | 58.24 |
| CRF+RR | 52.10 | 78.20 | 70.15 | 78.97 | 7.08 | 9.91 | 60.14 | **59.74** | 60.80 |
| TextCNN | 81.12 | 68.92 | 71.42 | 79.81 | 10.12 | 11.51 | 63.51 | 44.76 | 51.62 |
| [39] | 70.78 | 58.42 | 65.65 | 59.13 | 1.71 | 6.78 | 70.12 | 51.78 | 57.45 |
| Ours (C) | **89.61** | **100** | **94.52** | **100** | 3.48 | 6.72 | 63.39 | 43.56 | 51.64 |

| Ours (S) | 100 | 57.04 | **72.64** | 100 | 15.38 | 26.67 | 99.11 | 47.44 | **64.16** |
| C+S | 100 | 57.04 | **72.64** | 100 | 1.73 | 3.4 | 99.11 | 47.44 | **64.16** |

# 5 Conclusion

Early detection of online rumor posts is of utmost importance. We have introduced an unsupervised mechanism for distinguishing rumors from non-rumor tweets. The proposed methodology leverages the content and social features of online posts and surpasses the performances of several supervised state-of-the-art algorithms. From the results obtained, it is evident that features from a social media post are correlated and can be amalgamated using dimensionality reduction to perform unsupervised clustering yielding high accuracy. This work endeavors to delimit the challenges of supervised learning, which requires large datasets and long training durations, which is hardly possible in online scenarios. Our unsupervised framework paves the way for advancements in early rumor identification to classify posts as soon as they appear online. We validate our methodology on nine events of the PHEME dataset. Results demonstrate that the combination of content and social features is eminent for rumor identification. Among the clustering algorithms RBF spectral clustering and Fuzzy C means performed at par with NN spectral clustering, whereas K-means demonstrated a mediocre performance. Compared with existing baselines, the proposed methodology demonstrates ~25-30% improvement in several datasets.

Since the development in unsupervised and semi-supervised algorithms is low paced in the rumor detection domain, it opens multiple future directions for fellow researchers to follow. We intend to reduce feature attributes used to characterize rumors and rumors to perform the detection by the least number of attributes. Also, we plan to perform fine-grained classification to identify further the stance of posts identified as rumors. The proposed methodology can be strengthened by optimizing the clustering algorithms to generalize social media attributes. Therefore, further attempts are required in the direction of early rumor detection to verify online posts dynamically.

# References

[1] A. Zubiaga, H. Ji, and K. Knight, "Curating and contextualizing twitter stories to assist with social newsgathering," 2013. doi: 10.1145/2449396.2449424.

[2] N. Diakopoulos, M. de Choudhury, and M. Naaman, "Finding and assessing social media information sources in the context of journalism," 2012. doi: 10.1145/2207676.2208409.

[3] P. Tolmie, R. Procter, M. Rouncefield, M. Liakata, and A. Zubiaga, "Microblog Analysis as a Program of Work," *ACM Transactions on Social Computing*, vol. 1, no. 1, 2018, doi: 10.1145/3162956.

[4] H. Kwak, C. Lee, H. Park, and S. Moon, "What is Twitter, a social network or a news media?," 2010. doi: 10.1145/1772690.1772751.

[5] L. Y. C. Wong and J. Burkell, "Motivations for sharing news on social media," in *ACM International Conference Proceeding Series*, 2017, vol. Part F129683. doi: 10.1145/3097286.3097343.

[6] N. Bai, F. Meng, X. Rui, and Z. Wang, "Rumour Detection Based on Graph Convolutional Neural Net," *IEEE Access*, vol. 9, 2021, doi: 10.1109/ACCESS.2021.3050563.

[7] N. Tam, "Graph-based rumour detection for social media," 2019.

[8] G. Cai, H. Wu, and R. Lv, "Rumors detection in Chinese via crowd responses," 2014. doi: 10.1109/ASONAM.2014.6921694.

[9] N. DiFonzo and P. Bordia, "Rumor, gossip and urban legends," *Diogenes*, vol. 54, no. 1. 2007. doi: 10.1177/0392192107073433.

[10] S. Vosoughi, M. 'Neo' Mohsenvand, and D. Roy, "Rumor gauge: Predicting the veracity of rumors on twitter," *ACM Transactions on Knowledge Discovery from Data*, vol. 11, no. 4, 2017, doi: 10.1145/3070644.

[11] Z. Zhao, P. Resnick, and Q. Mei, "Enquiring minds: Early detection of rumors in social media from enquiry posts," 2015. doi: 10.1145/2736277.2741637.

[12] P. Tolmie *et al.*, "Supporting the use of user generated content in journalistic practice," in *Conference on Human Factors in Computing Systems - Proceedings*, 2017, vol. 2017-May. doi: 10.1145/3025453.3025892.

[13] U. Bhattacharjee, P. K. Srijith, and M. S. Desarkar, "Term Specific TF-IDF Boosting for Detection of Rumours in Social Networks," 2019. doi: 10.1109/COMSNETS.2019.8711427.

[14] R. McCreadie, C. Macdonald, and I. Ounis, "Crowdsourced rumour identification during emergencies," 2015. doi: 10.1145/2740908.2742573.

[15] A. Bondielli and F. Marcelloni, "A survey on fake news and rumour detection techniques," *Information Sciences*, vol. 497, 2019, doi: 10.1016/j.ins.2019.05.035.

[16] D. Varshney and D. K. Vishwakarma, "A review on rumour prediction and veracity assessment in online social network," *Expert Systems with Applications*, vol. 168. 2021. doi: 10.1016/j.eswa.2020.114208.

[17] V. Qazvinian, E. Rosengren, D. R. Radev, and Q. Mei, "Rumor has it: Identifying misinformation in microblogs," 2011.

[18] C. Castillo, M. Mendoza, and B. Poblete, "Information credibility on Twitter," 2011. doi: 10.1145/1963405.1963500.

[19] S. Hamidian and M. Diab, "Rumor Identification and Belief Investigation on Twitter," 2016. doi: 10.18653/v1/w16-0403.

[20] P. Meel and D. K. Vishwakarma, "Fake news, rumor, information pollution in social media and web: A contemporary survey of state-of-the-arts, challenges and opportunities," *Expert Systems with Applications*, vol. 153. 2020. doi: 10.1016/j.eswa.2019.112986.

[21] A. Zubiaga, M. Liakata, and R. Procter, "Exploiting context for rumour detection in social media," in *Lecture Notes in Computer Science (including subseries Lecture Notes in Artificial Intelligence and Lecture Notes in Bioinformatics)*, 2017, vol. 10539 LNCS. doi: 10.1007/978-3-319-67217-5_8.

[22] S. Kwon, M. Cha, and K. Jung, "Rumor detection over varying time windows," *PLoS ONE*, vol. 12, no. 1, 2017, doi: 10.1371/journal.pone.0168344.

[23] S. Kwon, M. Cha, K. Jung, W. Chen, and Y. Wang, "Prominent features of rumor propagation in online social media," 2013. doi: 10.1109/ICDM.2013.61.

[24] E. Agichtein, C. Castillo, D. Donato, A. Gionis, and G. Mishne, "Finding high-quality content in social media," 2008. doi: 10.1145/1341531.1341557.

[25] O. Alonso, C. Carson, and D. Gerster, "Detecting uninteresting content in text streams," *… for Search Evaluation …*, no. Cse, 2010.

[26] A. Lee Hughes and L. Palen, "Twitter adoption and use in mass convergence and emergency events," *International Journal of Emergency Management*, vol. 6, no. 3–4, 2009, doi: 10.1504/IJEM.2009.031564.

[27] A. Java, X. Song, T. Finin, and B. Tseng, "Why we twitter: Understanding microblogging usage and communities," 2007. doi: 10.1145/1348549.1348556.

[28] F. Yang, X. Yu, Y. Liu, and M. Yang, "Automatic detection of rumor on Sina Weibo," 2012. doi: 10.1145/2350190.2350203.

[29] J. Ma, W. Gao, Z. Wei, Y. Lu, and K. F. Wong, "Detect rumors using time series of social context information on microblogging websites," in *International Conference on Information and Knowledge Management, Proceedings*, 2015, vol. 19-23-Oct-2015. doi: 10.1145/2806416.2806607.

[30] J. Sampson, F. Morstatter, L. Wu, and H. Liu, "Leveraging the implicit structure within social media for emergent rumor detection," in *International Conference on Information and Knowledge Management, Proceedings*, 2016, vol. 24-28-October-2016. doi: 10.1145/2983323.2983697.

[31] K. Wu, S. Yang, and K. Q. Zhu, "False rumors detection on Sina Weibo by propagation structures," in *Proceedings - International Conference on Data Engineering*, 2015, vol. 2015-May. doi: 10.1109/ICDE.2015.7113322.

[32] Z. Jin, J. Cao, Y. Zhang, and Y. Zhang, "MCG-ICT at MediaEval 2015: Verifying multimedia use with a two-level classification model," in *CEUR Workshop Proceedings*, 2015, vol. 1436.

[33] A. Ebrahimi Fard, M. Mohammadi, Y. Chen, and B. van de Walle, "Computational Rumor Detection Without Non-Rumor: A One-Class Classification Approach," *IEEE Transactions on Computational Social Systems*, vol. 6, no. 5, 2019, doi: 10.1109/TCSS.2019.2931186.

[34] Y. Geng, J. Sui, and Q. Zhu, "Rumor detection of Sina Weibo based on SDSMOTE and feature selection," 2019. doi: 10.1109/ICCCBDA.2019.8725715.

[35] S. Santhoshkumar and L. D. Dhinesh Babu, "Earlier detection of rumors in online social networks using certainty-factor-based convolutional neural networks," *Social Network Analysis and Mining*, vol. 10, no. 1, 2020, doi: 10.1007/s13278-020-00634-x.

[36] C. M. M. Kotteti, X. Dong, and L. Qian, "Ensemble deep learning on time-series representation of tweets for rumor detection in social media," *Applied Sciences (Switzerland)*, vol. 10, no. 21, 2020, doi: 10.3390/app10217541.

[37] Z. Jin, J. Cao, H. Guo, Y. Zhang, and J. Luo, "Multimodal fusion with recurrent neural networks for rumor detection on microblogs," 2017. doi: 10.1145/3123266.3123454.

[38] S. Lathiya, J. S. Dhobi, A. Zubiaga, M. Liakata, and R. Procter, "Birds of a feather check together: Leveraging homophily for sequential rumour detection," *Online Social Networks and Media*, vol. 19, 2020, doi: 10.1016/j.osnem.2020.100097.

[39] X. Liu, A. Nourbakhsh, Q. Li, R. Fang, and S. Shah, "Real-time rumor debunking on twitter," in *International Conference on Information and Knowledge Management, Proceedings*, 2015, vol. 19-23-Oct-2015. doi: 10.1145/2806416.2806651.